# DAMPING FACTORS FOR HEAD-TAIL MODES AT STRONG SPACE CHARGE

Alexey Burov, Fermilab, Batavia, IL 60510

*Abstract*

This paper suggests how feedback and Landau damping can be taken into account for transverse oscillations of bunched beam at strong space charge.

## MAIN EQUATION

Space charge is known to be able to change dramatically collective modes [1-3]. For transverse oscillations of bunched beams, a parameter of the space charge strength is a ratio of the maximal space charge tune shift to the synchrotron tune. When this parameter is large, the transverse oscillations are described by a one-dimensional integro-differential equation derived in Ref. [2] and reproduced here for the sake of convenience:

$$\nu y + \frac{1}{Q_{\text{eff}}} \frac{d}{d\tau}\left(u^2 \frac{dy}{d\tau}\right) = \kappa N \left(\hat{W}y + Dy\right)$$

$$\hat{W}y = \int_{-\infty}^{\infty} W(\tau - s)\exp\left[i\zeta(\tau - s)\right]\rho(s)y(s)ds$$

$$Dy = y(\tau)\int_{-\infty}^{\infty} D(\tau - s)\rho(s)ds$$

$$\left.\frac{dy}{d\tau}\right|_{\tau \to \pm\infty} = 0$$
(1)

Here $y = y(\tau)$ and $\nu$ are the eigenfunction and the eigenvalue to be found for the bunch transverse oscillations, $\tau$ and $s$ are longitudinal positions within the bunch, $W$ and $D$ are the dipole and quadrupole (or the driving and detuning) wakes, $\rho$ is the normalized line density

$$\int \rho\, ds = 1 \;, \tag{2}$$

$N$ is the number of particles per bunch,

$$\kappa = \frac{r_0 R}{4\pi \beta^2 \gamma Q_b} \;, \tag{3}$$

with $r_0$ as the classical radius, $R$ as the average machine radius, $\beta$ and $\gamma$ as the relativistic factors and $Q_b$ as the bare betatron tune. The parameter $\zeta$ staying in the exponents of the wake integral is a negated ratio of the chromaticity $\xi = p\, dQ_b/dp$ and the slippage factor $\eta = \gamma_t^{-2} - \gamma^{-2}$, i.e. $\zeta = -\xi/\eta$. The symbol $Q_{\text{eff}} = Q_{\text{eff}}(\tau)$ stays for the space charge tune shift at the given position along the bunch $\tau$, averaged over the both transverse action, see Ref. [2]. Thus, the effective space charge tune shift is proportional to the local line density:

$$Q_{\text{eff}}(\tau) = Q_{\text{eff}}(0)\rho(\tau)/\rho(0) \;. \tag{4}$$

For the transversely Gaussian bunch,

$$Q_{\text{eff}}(\tau) = 0.52\, Q_{\max}(\tau) \tag{5}$$

where $Q_{\max}(\tau)$ is the space charge tune shift at the bunch axis. The symbol $u^2 = u^2(\tau)$ stays for the average square of the particle longitudinal velocity $v_i = d\tau_i/d\theta$, with time measured as the angle $\theta$ along the machine circumference, at the given position $\tau$:

$$u^2 = \langle v^2 \rangle = \frac{\int f(v,\tau) v^2\, dv}{\int f(v,\tau)\, dv} \;, \tag{6}$$

where $f(v,\tau)$ is the longitudinal distribution function. For the longitudinally Gaussian distribution with the rms bunch length $\sigma_\tau$, the temperature function $u^2$ is constant along the bunch:

$$u^2 = Q_s^2 \sigma_\tau^2 \;, \tag{7}$$

where $Q_s$ is the synchrotron tune.

In general, the wake term $\hat{W}y$ is a sum of single-bunch (SB), coupled-bunch (CB) wakes and the damper (G) terms:

$$\hat{W}y = \hat{W}_{\text{SB}}y + \hat{W}_{\text{CB}}y + \hat{G}y \tag{8}$$

The single-bunch term $\hat{W}_{\text{SB}}y$ is described exactly as in Eq. (1), where only $s > \tau$ contributes due to the causality, and the integral is taken along the single-bunch interval only:

$$\hat{W}_{\text{SB}}y = \int_{\text{SB}} W(\tau - s)\exp\left[i\zeta(\tau - s)\right]\rho(s)y(s)ds \tag{9}$$

The coupled-bunch term results from summations of the fields left by preceding passages of the bunches through the given position of the ring. This summation is especially simple when the bunches are equidistant. In this case, due to the symmetry, the offsets of the neighbor bunches, being taken at the same time, differ only by the phase factors $\exp(i\psi_\mu)$:

$$y(s + s_0) = \exp(i\psi_\mu)y(s) \;. \tag{10}$$

For $M$ bunches in the ring,

$$\psi_\mu = \frac{2\pi\mu}{M}; \quad \mu = 0,1,\ldots,M-1, \quad (11)$$

where integer $\mu$ is a counter of the coupled-bunch modes. After that, we have to take into account that the given reference bunch sees the fields left behind by other bunches not at the same time, but certain time ago, proportional to the distance between the bunches. This leads to an additional time-related factor to be taken into account together with the space-related phase factor:

$$y(s+s_0, \theta-s_0) = \exp(i\phi_\mu) y(s,\theta);$$
$$\phi_\mu = \psi_\mu + \frac{2\pi Q_b}{M}. \quad (12)$$

Remember that both time and space are measured as the angles of revolution, and the leading particles have higher coordinate than the following ones. From here, the coupled-bunch contribution in Eq. (8) follows as a single-bunch integral:

$$\hat{W}_{CB} y = \int_{SB} \tilde{W}_\mu(\tau,s) \exp[i\zeta(\tau-s)] \rho(s) y(s) ds;$$
$$\tilde{W}_\mu(\tau,s) = \sum_{k=1}^\infty W(\tau-s-ks_0) \exp(ik\phi_\mu). \quad (13)$$

For those cases when the wake function of the neighbour bunch does not change much along the reference bunch, i.e. the coupled-bunch wake is flat [4],

$$W(\tau-s-ks_0) \approx W(-ks_0) \quad (14)$$

the result of summation in the right hand side of Eq. (13) does not depend on the specific positions $s$, $\tau$ within the bunches; thus, the effective coupled bunch wake $\tilde{W}_\mu$ is a constant which can be taken out of the integral:

$$\hat{W}_{CB} y = \tilde{W}_\mu e^{i\zeta\tau} \int_{SB} e^{-i\zeta s} \rho(s) y(s) ds. \quad (15)$$

In principle, the damper term $\hat{G}y$ in Eq. (8) is similar to the coupled-bunch one. If the feedback bandwidth is much smaller than the inverse bunch length, the damper takes just one parameter per bunch. This parameter can be chosen as an offset of the centre of mass, and the kick can be designed to be flat along the bunch. Then the damper term is represented similar to Eq. (15):

$$\hat{G}y = \tilde{G}_\mu e^{i\zeta\tau} \int_{SB} e^{-i\zeta s} \rho(s) y(s) ds, \quad (16)$$

If the damper is bunch-by-bunch, there is no coupled-bunch mode dependence in the feedback factor $\tilde{G}_\mu$, i.e. $\tilde{G}_\mu = \tilde{G}$.

Similarly to the driving wake factor, Eq. (8), there is a certain coupled-bunch contribution in the detuning wake as well:

$$Dy = D_{SB} y + D_{CB} y;$$
$$D_{SB} = \int_{SB} \rho(s) D(\tau-s) ds;$$
$$D_{CB} = \int_{SB} \rho(s) \tilde{D}(\tau,s) ds; \quad (17)$$
$$\tilde{D}(\tau,s) = \sum_{k=1}^\infty D(\tau-s-ks_0).$$

If the detuning wake is flat in the same sense as in Eq. (14), its coupled-bunch contribution is identical for all the particles, so it works as a constant quadrupole without any influence to the beam dynamics unless it leads to a dangerous resonance crossing.

## SOLUTION

To find the spectrum of Eq. (1), its eigenfunctions can be expanded over its zero-wake solutions $y_k^0(\tau)$ satisfying the following equation:

$$\nu^0 y^0 + \frac{1}{Q_{eff}} \frac{d}{d\tau}\left(u^2 \frac{dy^0}{d\tau}\right) = 0;$$
$$\left.\frac{dy^0}{d\tau}\right|_{\tau\to\pm\infty} = 0 \quad (18)$$

All the eigenfunctions $y_k^0(\tau)$ are orthogonal and can be normalized, so that:

$$\int_{SB} ds \rho(s) y_k^0(s) y_m^0(s) = \delta_{km}. \quad (19)$$

For the Gaussian distribution, the spectrum of this equation has been described in Ref.[2,3]; similarly, it can be found for any potential well and distribution function.

Expansion of the eigenfunction $y(\tau)$ over the no-wake set $y^0(\tau)$,

$$y(\tau) = \sum_{k=0}^\infty B_k y_k^0(\tau) \quad (20)$$

with the amplitudes $B$ to be found, with the following multiplication of Eq.(1) on $y_l^0(\tau)\rho(\tau)$ and its integration over the bunch length, leads to:

$$\left[\kappa N\hat{\mathbf{W}} + \kappa N\hat{\mathbf{D}} + \text{Diag}(\nu^0)\right]\mathbf{B} = \nu\mathbf{B}. \quad (21)$$

Here $\hat{\mathbf{W}}$ and $\hat{\mathbf{D}}$ are the matrices of the driving and detuning wakes in the basis of the no-wake modes of Eq. (18), (19):

$$\left(\hat{\mathbf{W}}_{\text{SB}}\right)_{lk} = \iint_{\text{SB}} W(\tau-s)\rho(\tau)\rho(s)e^{i\zeta(\tau-s)}y_l^0(\tau)y_k^0(s)d\tau ds;$$
$$\left(\hat{\mathbf{W}}_{\text{CB}}\right)_{lk} = \iint_{\text{SB}} \tilde{W}_\mu(\tau,s)\rho(\tau)\rho(s)e^{i\zeta(\tau-s)}y_l^0(\tau)y_k^0(s)d\tau ds; \quad (22)$$

$$\hat{\mathbf{G}}_{lk} = \tilde{G}_\mu I_l(\zeta)I_k^*(\zeta);$$
$$I_l(\zeta) = \int_{\text{SB}} \rho(\tau)e^{i\zeta\tau}y_l^0(\tau)d\tau; \quad (23)$$

$$\left(\hat{\mathbf{D}}_{\text{SB}}\right)_{lk} = \int_{\text{SB}} F_{\text{SB}}(\tau)\rho(\tau)y_l^0(\tau)y_k^0(\tau)d\tau;$$
$$F_{\text{SB}}(\tau) = \int_{\text{SB}} D(\tau-s)\rho(s)ds; \quad (24)$$

$$\left(\hat{\mathbf{D}}_{\text{CB}}\right)_{lk} = \int_{\text{SB}} F_{\text{CB}}(\tau)\rho(\tau)y_l^0(\tau)y_k^0(\tau)d\tau;$$
$$F_{\text{CB}}(\tau) = \int_{\text{SB}} \tilde{D}(\tau,s)\rho(s)ds. \quad (25)$$

## DAMPER DETAILS

In case when the feedback takes something different from the centre of mass and its kick is not flat over the bunch, the damper matrix has to be modified with provided pickup and kicker functions $P(s), K(\tau)$:

$$\hat{\mathbf{G}}_{lk} = \tilde{G}_\mu K_l(\zeta)P_k^*(\zeta);$$
$$K_l(\zeta) = \int_{\text{SB}} K(\tau)\rho(\tau)e^{i\zeta\tau}y_l^0(\tau)d\tau; \quad (26)$$
$$P_k^*(\zeta) = \int_{\text{SB}} P(s)\rho(s)e^{-i\zeta s}y_l^0(s)ds.$$

Equation (21) is a standard linear algebra eigensystem problem which solution is straightforward as soon as the wake functions, the feedback properties, the potential well, and the beam distribution functions, longitudinal and transverse, are given. This equation allows computing the instability growth rates for fairly general situations when the Landau damping can be neglected. However, without Landau damping nothing can be said about the instability threshold, so the theory is significantly incomplete.

## INSTABILITY THRESHOLDS

For the strong space charge, Landau damping rates were roughly estimated in Ref. [2]. Numerical simulations give a possibility for more accurate knowledge of the damping rates, with the numerical factors to be found with a good precision. This work has been started by V. Kornilov and O. Boine-Frankenheim several years ago with their PATRIC code [5,6], has been joined recently by A. Macridin et al. with the Synergia program [7]. As soon as the Landau damping rates are reliably established, they can be introduced in Eq. (21) as externally given imaginary parts of the vector $\nu^0$ in Eq. (21) with the following substitution:

$$\nu_k^0 \to \nu_k^0 - i\lambda_k, \quad (27)$$

where $\lambda_k$ is Landau damping rate for the no-wake eigenfunction $y_k^0(\tau)$. When the collective tune shifts imposed by the wakes are small compared to the synchrotron tune, their influence on the Landau damping can be neglected. Potential importance of the image charges and currents for Landau damping was shown in Ref. [5,6]. As soon as Landau damping rates are included in Eq. (21), theory of transverse stability of bunched beams with strong space charge would be complete. Hopefully, this work will be done in a reasonable future.